\documentclass[12pt]{article}
\usepackage{amsmath,amssymb,epsfig}
\makeatletter \@addtoreset{equation}{section} \makeatother
\renewcommand{\theequation}{\thesection.\arabic{equation}}
\newcommand{\ba}{\begin{array}}
\newcommand{\ea}{\end{array}}
\newcommand{\beq}{\begin{equation}}
\newcommand{\eeq}{\end{equation}}
\newcommand{\bea}{\begin{eqnarray}}
\newcommand{\eea}{\end{eqnarray}}
\def\bce{\begin{center}}
\def\ece{\end{center}}

\def\nonu{\nonumber}

\def\pa{\partial}

\def\be{\beta}

\def\de{\delta}

\def\la{\lambda}

\def\diag{\mathop{\rm diag}}

\def\eps6{{\displaystyle \mathop{\epsilon}^{6}}{}}

\def\nab6{{\displaystyle \mathop{\nabla}^{6}}{}}
\def\0{{\sst{(0)}}}
\def\1{{\sst{(1)}}}
\def\2{{\sst{(2)}}}
\def\3{{\sst{(3)}}}
\def\4{{\sst{(4)}}}
\def\5{{\sst{(5)}}}
\def\6{{\sst{(6)}}}
\def\7{{\sst{(7)}}}
\def\8{{\sst{(8)}}}

\def\ba{\begin{array}}
\def\ea{\end{array}}
\def\beq{\begin{equation}}
\def\eeq{\end{equation}}
\def\be{\begin{equation}}
\def\ee{\end{equation}}

\def\Tr{\mathop{\rm Tr}}

\def\diag{\mathop{\rm diag}}

\def\la{\lambda}
\def\eps{\epsilon}

\def\ba{\begin{array}}
\def\ea{\end{array}}
\def\beq{\begin{equation}}
\def\eeq{\end{equation}}
\def\be{\begin{equation}}
\def\ee{\end{equation}}

\def\Tr{\mathop{\rm Tr}}

\def\diag{\mathop{\rm diag}}

\def\la{\lambda}
\def\eps{\epsilon}

\newcommand{\bean}{\begin{eqnarray*}}
\newcommand{\eean}{\end{eqnarray*}}

\begin{document}
\thispagestyle{empty} \addtocounter{page}{-1}
   \begin{flushright}
%KIAS-P08nnn \\
%CALT-68-nnnn \\
%{\tt hep-th/yymmnnn}\\
\end{flushright}

\vspace*{1.3cm}

 \centerline{ \Large \bf  Holographic Supergravity Dual to } 
\vspace{.3cm} 
\centerline{ \Large \bf  
Three Dimensional ${\cal N}=2$ Gauge Theory   } 
\vspace*{1.5cm}
\centerline{{\bf Changhyun Ahn }
%, {\bf Kazuo Hosomichi $^{2}$}
%and {\bf Sungjay Lee $^{2}$} 
} 
\vspace*{1.0cm} 
\centerline{\it  
Department of Physics, Kyungpook National University, Taegu
702-701, Korea} 
%\centerline{\it $^{2}$ Korea Institute for 
%Advanced Study, Seoul 130-012, Korea }
\vspace*{0.8cm} 
\centerline{\tt ahn@knu.ac.kr
%, \qquad
%hosomiti@kias.re.kr, \qquad sjlee@kias.re.kr
} 
\vskip2cm

\centerline{\bf Abstract}
\vspace*{0.5cm}

By examining the previously known holographic ${\cal N}=2$
supersymmetric renormalization group flow solution in four dimensions, 
we describe the mass-deformed Bagger-Lambert theory, 
that has $SU(3)_I \times U(1)_R$
symmetry, by the addition of mass term
for one of the four adjoint chiral superfields as its dual theory. 
A further detailed correspondence between fields of
$AdS_4$ supergravity and composite operators of the 
infrared field theory is obtained.
 
\baselineskip=18pt
\newpage
\renewcommand{\theequation}
{\arabic{section}\mbox{.}\arabic{equation}}

%%%%%%%%%%%%%%%%%%%%%%%%%%%%%%%%%%%%%%%%%%%%%%%%%%%%%%%%%%%%%%%%%%%%%
%%%%%%%%%%%%%%%%%%%%%%%%%%%%%%%%%%%%%%%%%%%%%%%%%%%%%%%%%%%%%%%%%%%%%%
\section{Introduction}
%%%%%%%%%%%%%%%%%%%%%%%%%%%%%%%%%%%%%%%%%%%%%%%%%%%%%%%%%%%%%%%%%%%%%%
%%%%%%%%%%%%%%%%%%%%%%%%%%%%%%%%%%%%%%%%%%%%%%%%%%%%%%%%%%%%%%%%%%%%%

The holographic theory on M2-branes is given by 
an ${\cal N}=8$ supersymmetric theory with eight scalars, eight
fermions and sixteen supercharges. 
The $AdS_4 \times {\bf S}^7$ background yields 
the holographic dual of a strongly coupled superconformal fixed point
\cite{Maldacena,Seiberg}.
By lifting the renormalization group(RG) flow in four dimensions \cite{AP,AW}
to eleven dimensions,
the M-theory solutions \cite{CPW} which are holographic duals of flows of the
maximally supersymmetric theory in three dimensions are examined. 
Giving one of four complex superfields a mass leads to an ${\cal N}=2$
supersymmetric flow(four supersymmetries) to a new superconformal
fixed point. The vacuum expectation values 
of remaining three complex superfields parametrize the Coulomb branch 
at this fixed point.  
A M2-brane probe analysis of the supergravity solution 
shows a three complex-dimensional space of moduli  for 
the brane probe \cite{JLP01}. 
However, the microscopic configuration of coincident M2-branes was
still lacking.
 
Recently, Bagger and Lambert(BL) have proposed a Lagrangian to describe 
the low energy dynamics of coincident M2-branes in \cite{BL0711}.
See also relevant papers \cite{BL06,BL0712,Gustavsson07,Gustavsson08}.
This BL theory 
is three dimensional ${\cal N}=8$ supersymmetric field
theory with $SO(8)$ global symmetry 
based on new algebraic structure, 3-algebra.
In particular, 3-algebra with Lorentzian signature was proposed
by \cite{GMR,BRTV,HIM}.
The generators of the 3-algebra are the generators of an arbitrary
semisimple Lie algebra plus two additional null generators.
This theory with gauge group 
$SU(N)$ is a good candidate for the theory of $N$ coincident M2-branes.
We list some relevant works on the BL theory, from various different
point of views, in \cite{ABJM}-\cite{MP}. 

In this paper, 
starting from the first order differential equations, that are the
supersymmetric flow solution in four dimensional ${\cal N}=8$ gauged
supergravity
interpolating between an exterior $AdS_4$ region with maximal
supersymmetry
and an interior $AdS_4$ with one quarter of the maximal supersymmetry,  
we want to interpret this as the RG flow in ${\cal N}=8$ BL theory
which has $OSp(8|4)$ symmetry
broken to an ${\cal N}=2$ theory which has $OSp(2|4)$ symmetry
by the addition of a mass term for
one of the four adjoint chiral superfields. A precise correspondence 
is obtained between fields of bulk supergravity in the $AdS_4$ region
and composite operators of the IR field theory in three dimensions 
\footnote{For the $AdS_5 \times {\bf S}^5$ background with D3-branes, it is
well known in \cite{FGPW} that the holographic dual is studied for the flow
of ${\cal N}=4$ super Yang-Mills theory to the ${\cal N}=1$
supersymmetric Leigh-Strassler fixed point \cite{LS}. There exist
earlier works on the M-theory flow solutions in 11 dimensions 
\cite{NW04}. }. 
Since the Lagrangian is known, one can check how the supersymmetry
breaks for specific deformation and can extract the correct full
superpotential including the superpotential before the deformation 
also. 
One would like to see the three dimensional analog of Leigh-Strassler 
\cite{LS}
RG flow in mass-deformed BL theory in three dimensions
by looking at its holographic dual theory in four dimensions along the
line of \cite{CPW}.   

In section 2, we review the supergravity solution in four dimensions
in the context of RG flow, describe two supergravity critical points 
and present the supergravity multiplet in terms of 
$SU(3)_I \times U(1)_Y$ invariant ones \footnote{We put the index $I$
in $SU(3)$ group for ``invariance'' 
in order to emphasize that along the flow $SU(3)$
group is preserved. The index $Y$ in $U(1)_Y$ is for the hypercharge
in the context of $AdS_4$ supergravity and is related to $U(1)_R$
charge in the context of boundary gauge theory.}.

In section 3, we deform BL theory by adding one of the mass term among
four chiral superfields, along the lines of \cite{HLL,GSP}, write down
the superpotential in ${\cal N}=2$ superfields and describe the scale
dimensions for the superfields at UV and IR.  

In section 4, 
the $OSp(2|4)$ representations(energy, spin, hypercharge) and 
$SU(3)_I$ representations
in the supergravity mass spectrum for
each multiplet at
the ${\cal N}=2$ critical point and the corresponding ${\cal N}=2$
superfield in the boundary gauge theory are given.
The Kahler potential at IR is obtained.

In section 5, 
we end up with the future directions.

%%%%%%%%%%%%%%%%%%%%%%%%%%%%%%%%%%%%%%%%%%%%%%%%%%%%%%%%%%%%%%%%
%%%%%%%%%%%%%%%%%%%%%%%%%%%%%%%%%%%%%%%%%%%%%%%%%%%%%%%%%%%%%%%%
\section{The holographic ${\cal N}=2$ 
supersymmetric RG flow in four dimensions }
%%%%%%%%%%%%%%%%%%%%%%%%%%%%%%%%%%%%%%%%%%%%%%%%%%%%%%%%%%%%%%%%
%%%%%%%%%%%%%%%%%%%%%%%%%%%%%%%%%%%%%%%%%%%%%%%%%%%%%%%%%%%%%%%%

By gauging the $SO(8)$ subgroup of $E_7$ in the global $E_7$ $\times$
local $SU(8)$ supergravity \cite{CJS},
de Wit and Nicolai \cite{dN82} 
constructed a four-dimensional supergravity theory. 
This theory has self-interaction of a single
massless ${\cal N}=8$ supermultiplet of spins 
$(2, \frac{3}{2}, 1, \frac{1}{2}, 0^{\pm})$ but with 
local $SO(8)$ $\times$ local
$SU(8)$ invariance. 
It is well known \cite{CJ78} that the 70 real 
scalars of ${\cal N}=8$ supergravity
live on the coset space $E_{7(7)}/SU(8)$ because 63 fields may be gauged
away by an $SU(8)$ rotation and are described by 
an element of the fundamental 56-dimensional representation
of $E_7$.
Then the effective nontrivial
potential arising from $SO(8)$ gauging can be written 
in compact form.
The complex self-dual tensor describes the 35 
scalars 
and 35 pseudo-scalar fields  of ${\cal N}=8$ supergravity.
After gauge fixing, one does not distinguish between $SO(8)$ and $SU(8)$
indices. 
The full supersymmetric solution where both 
scalars and pseudo-scalars
vanish yields $SO(8)$ vacuum state with ${\cal N}=8$ supersymmetry. Note 
that $SU(8)$ is not a symmetry of the vacuum. 

It is known that, in ${\cal N}=8$ supergravity, there also exists 
a ${\cal N}=2$ supersymmetric, $SU(3)_I \times U(1)_Y$ invariant vacuum \cite{NW}. 
To reach this critical point, one has to turn on expectation values of both 
scalar $\la$ and pseudo-scalar $\la'$ fields 
where the 
completely antisymmetric self-dual and anti-self-dual
tensors are invariant under $SU(3)_I \times U(1)_Y$.
Therefore 56-beins can be written as $ 56\times 56$ matrix whose
elements are some functions of scalar and pseudo-scalars. 
Then the $SU(3)_I \times U(1)_Y$-invariant 
scalar potential of ${\cal N}=8$ supergravity is given by 
\cite{Warner83,NW,AP}
\bea
V(\la, \la')= g^2 \left[ \frac{16}{3}  \left(\frac{\partial W}{\partial 
\la} \right)^2 + 4   \left(\frac{\partial W}{\partial 
\la'} \right)^2 - 6  W^2 \right]
\label{pot}
\eea
where $g$ is $SO(8)$ gauge coupling constant and  
the superpotential can be written as \cite{AP,CPW} 
\bea
W(\la, \la')  =  \frac{1}{16} e^{-\frac{1}{2\sqrt{2}} \la -\sqrt{2} \la'}
 \left( 3 - e^{\sqrt{2} \la} + 6 e^{\sqrt{2} \la'} +3
e^{2\sqrt{2} \la'}+6 e^{\sqrt{2} ( \la + \la')} -e^{
\sqrt{2} (\la +2 \la')} \right).
\label{W}
\eea

There are two critical points and we summarize these in Table 1.

%%%%%%%%%%%%%%%%%%%%%%%%%%%%%%%%%%%%%%%%%
$\bullet$ $SO(8)$ critical point
%%%%%%%%%%%%%%%%%%%%%%%%%%%%%%%%%%%%%%%%%%

There is well-known, trivial critical point at which all the scalars
vanish($\lambda=\lambda'=0$) 
and whose cosmological constant $\Lambda=-6g^2$ from (\ref{pot}) and which 
preserves ${\cal N}=8$ supersymmetry. 

%%%%%%%%%%%%%%%%%%%%%%%%%%%%%%%%%%%%%%%%%%%%%%%%
$\bullet$ $SU(3)_I \times U(1)_Y$ critical point
%%%%%%%%%%%%%%%%%%%%%%%%%%%%%%%%%%%%%%%%%%%%%%%%

There is a critical point at $\lambda = \sqrt{2} \sinh^{-1}
\left(\frac{1}{\sqrt{3}}\right)$
and $\lambda' = \sqrt{2} \sinh^{-1} \left(\frac{1}{\sqrt{2}}\right)$
and the cosmological constant $\Lambda=-\frac{9 \sqrt{3}}{2} g^2$.
This critical point has an unbroken ${\cal N}=2$ supersymmetry.

For the supergravity description of the 
nonconformal RG flow from one scale to 
another connecting the two critical points, 
the three dimensional Poincare invariant metric takes the form 
$
ds^2= e^{2A(r)} \eta_{\mu \nu} dx^{\mu} dx^{\nu} + dr^2$
where $\eta_{\mu \nu}=(-,+,+)$
and $r$ is the coordinate transverse to the domain wall.
Then the supersymmetric flow equations \cite{AP,CPW} with (\ref{W}) are 
described as
\bea
\frac{d \la}{d r}  =   
\frac{8}{3} \sqrt{2} g  \frac{\partial W}{\partial \la}, \qquad
\frac{d \la'}{d r}  =   
2 \sqrt{2} g  \frac{\partial W}{\partial \la'}, \qquad
\frac{d A}{d r}  =  - \sqrt{2} g W.
\label{solution}
\eea
The $AdS_4$ geometries at the end points imply conformal symmetry in
the UV and IR limits of the field theory.
We'll return to this when we discuss about the Kahler potential in
section 4.

%%%%%%%%%%%%%%%%%%%%%%%%%%%%%%%%%%%%%%%%%%%%%%%%%%%%%%%%%%%%%%%%%%%%%%
%table 1%%%%%%%%%%%%%%%%%%%%%%%%%%%%%%%%%%%%%%%%%%%%%%%%%%%%%%%%%%%%%%%%%%%%%
\begin{table} 
\begin{center}
\begin{tabular}{|c|c|c|c|c|} \hline
Symmetry & $\la$  &  $\la'$ & V & W
 \\ \hline
$SO(8)$ & $0$ & $0$ & $-6g^2$  &  $1$  \\
 $SU(3)_I \times U(1)_Y$ &  $\sqrt{2} \sinh^{-1} \frac{1}{\sqrt{3}}$ & 
$\sqrt{2} \sinh^{-1} \frac{1}{\sqrt{2}}$ & $-\frac{9 \sqrt{3}}{2} g^2$
 & $\frac{3^{\frac{3}{4}}}{2}$  \\
\hline 
\end{tabular} 
\end{center}
\caption{\sl 
Summary of two critical points with symmetry group, supergravity
fields, scalar potential and superpotential.}
%\label{tableso4}
\end{table} 
%%%%%%%%%%%%%%%%%%%%%%%%%%%%%%%%%%%%%%%%%%%%%%%%%%%%%%%%%%%%%%%%%%%%%%%%%%%%
%%%%%%%%%%%%%%%%%%%%%%%%%%%%%%%%%%%%%%%%%%%%%%%%%%%%%%%%%%%%%%%%%%%%%%%%%%%%

Since the unbroken group symmetry at the stationary point is 
$SU(3)_I \times U(1)_Y$, the fields of the ${\cal N}=8$ theory,
transforming in $SO(8)$ representations, should be 
decomposed into $SU(3)_I \times U(1)_Y$ representations. 
From the quadratic fermion terms of the gauged ${\cal N}=8$
supergravity Lagrangian \cite{NW},  
there exist the massless and massive graviton mass terms.
According to the decomposition $SO(8) 
\rightarrow SU(3)_I \times U(1)_Y$, 
the spin $\frac{3}{2}$ field breaks into \cite{NW}
\bea
{\bf 8} \rightarrow \left[{\bf 1}_{\frac{1}{2}} \oplus {\bf
  1}_{-\frac{1}{2}} \right] 
\oplus {\bf 3}_{\frac{1}{6}}
\oplus {\bf \bar{3}}_{-\frac{1}{6}},
\label{8}
\eea
and 
the two singlets in square bracket
correspond to the massless graviton of the ${\cal N}=2$ theory.
The other terms in 
the quadratic fermion terms of the gauged ${\cal N}=8$
supergravity Lagrangian provide
the spin $\frac{1}{2}$ masses and contain 
the Goldstino mass term. It turns out that
there is no octet term and so the octet mass term vanishes. 
The tensors in gauged ${\cal N}=8$ supergravity \cite{dN82}
have $SU(8)$ indices where the upper index transforms in ${\bf 8}$
and  the lower index transforms in ${\bf \bar{8}}$.
Using the charge normalization of \cite{NW}, one assigns 
the charges of $\frac{1}{6}$ and $-\frac{1}{6}$ to the lower indices 
$a$ and $\bar{a}$ respectively where $a =1,2,3$ and the lower indices 
$4$ and $\bar{4}$ should be assigned charges $\frac{1}{2}$ and 
$-\frac{1}{2}$ respectively. These charges appear in (\ref{8}).
A new complex basis is introduced with an index $A$ and $\bar{A}$
and the ${\bf 8}$ of $SO(8)$ in cartesian system is relabelled by 
$A$ and $\bar{A}$ where $A=a, 4$ and $\bar{A}=\bar{a}, \bar{4}$.

From the branching rule of $SO(8)$ into 
$SU(3)_I \times U(1)_Y$, the spin $\frac{1}{2}$ field transforms as \cite{NW}
\bea
{\bf 56} & \rightarrow & {\bf 1}_{\frac{1}{2}} \oplus
{\bf 1}_{-\frac{1}{2}} \oplus {\bf 6}_{-\frac{1}{6}} \oplus 
{\bf \bar{6}}_{\frac{1}{6}}  \oplus
{\bf 1}_{-\frac{1}{2}} \oplus
{\bf 1}_{\frac{1}{2}} \oplus \left[ {\bf 8}_{\frac{1}{2}} 
\oplus {\bf 8}_{-\frac{1}{2}} \right] \nonu \\
& \oplus & {\bf 3}_{\frac{1}{6}} \oplus
{\bf 3}_{-\frac{5}{6}} \oplus {\bf 3}_{\frac{1}{6}} \oplus 
{\bf \bar{3}}_{-\frac{1}{6}}  \oplus
{\bf \bar{3}}_{\frac{5}{6}} \oplus
{\bf \bar{3}}_{-\frac{1}{6}} \oplus \left[ {\bf 3}_{\frac{1}{6}} 
\oplus {\bf \bar{3}}_{-\frac{1}{6}} \right],  
\label{56}
\eea
and 
the six Goldstino modes that are absorbed into massive spin
$\frac{3}{2}$ fields are identified with triplets and anti-triplets in
square bracket and the two octets in square bracket
correspond to the massless vector multiplets of the ${\cal N}=2$ theory.
The decomposition of the vector fields with respect to $SO(8)$ \cite{NW}
\bea
{\bf 28} \rightarrow 
{\bf 1}_{0} \oplus {\bf 3}_{\frac{2}{3}} \oplus 
{\bf 3}_{-\frac{1}{3}}  \oplus
{\bf 3}_{-\frac{1}{3}} \oplus
{\bf \bar{3}}_{-\frac{2}{3}} \oplus 
{\bf \bar{3}}_{\frac{1}{3}} \oplus {\bf \bar{3}}_{\frac{1}{3}}
\oplus \left[{\bf 1}_{0}\right] 
\oplus \left[ {\bf 8}_{0} \right]
\label{28}
\eea
implies that the singlet in square bracket
corresponds to the massless graviton of the ${\cal N}=2$ theory while
the octet in square bracket
corresponds to the massless vector multiplets of the ${\cal N}=2$ theory.
Finally from the branching rule for spin 0 field \cite{NW}
\bea
{\bf 70} & \rightarrow &
{\bf 1}_{0} \oplus
{\bf 1}_{0} \oplus {\bf 1}_{1} \oplus 
{\bf 1}_{0}  \oplus
{\bf 1}_{-1} \oplus \left[ {\bf 8}_{0} 
\oplus {\bf 8}_{0} \right] 
\oplus  {\bf 3}_{-\frac{1}{3}} \oplus
{\bf \bar{3}}_{\frac{1}{3}} \oplus {\bf 6}_{\frac{1}{3}} \oplus 
{\bf 6}_{-\frac{2}{3}}  \oplus
{\bf \bar{6}}_{-\frac{1}{3}} \oplus
{\bf \bar{6}}_{\frac{2}{3}} \nonu \\
&\oplus& \left[ {\bf 1}_0 \oplus {\bf 3}_{\frac{2}{3}} 
\oplus {\bf 3}_{-\frac{1}{3}} \oplus {\bf 3}_{-\frac{1}{3}} \oplus
{\bf \bar{3}}_{-\frac{2}{3}} \oplus {\bf \bar{3}}_{\frac{1}{3}} \oplus
{\bf \bar{3}}_{\frac{1}{3}}  \right],
\label{70}
\eea
the two octets in square bracket
correspond to the massless vector multiplets of the ${\cal N}=2$
theory and the nineteen Goldstone bosons modes 
are identified with singlet, triplets and anti-triplets  in
square bracket. Their quantum numbers are in agreement with those of
massive vectors in (\ref{28}).

Finally, spin 2 field has the breaking ${\bf 1} \rightarrow {\bf 1}_0$
and is located at ${\cal N}=2$ massless graviton multiplet.

We'll rearrange (\ref{8}), (\ref{56}), (\ref{28}) and (\ref{70})
in the context of supergravity multiplet with corresponding $OSp(2|4)$
quantum numbers in section 4.  
The singlets are placed at long massive vector multiplet, triplets and
anti-triplets are located at short massive gravitino multiplet and 
sextets and anti-sextets sit in short massive hypermultiplet.

%%%%%%%%%%%%%%%%%%%%%%%%%%%%%%%%%%%%%%%%%%%%%%%%%%%%%%%%%%%%%%%%
%%%%%%%%%%%%%%%%%%%%%%%%%%%%%%%%%%%%%%%%%%%%%%%%%%%%%%%%%%%%%%%%
\section{An ${\cal N}=2$ supersymmetric membrane flow in three
  dimensional deformed BL theory }
%%%%%%%%%%%%%%%%%%%%%%%%%%%%%%%%%%%%%%%%%%%%%%%%%%%%%%%%%%%%%%%%
%%%%%%%%%%%%%%%%%%%%%%%%%%%%%%%%%%%%%%%%%%%%%%%%%%%%%%%%%%%%%%%%

The original BL Lagrangian \cite{BL0711} consists of 
the Chern-Simons terms, the kinetic terms for matter fields, the Yukawa
term and the potential term with supersymmetry
transformations on the gauge and matter fields.
The BF Lorentzian Lagrangian  \cite{GMR,BRTV,HIM} can be obtained 
by choosing structure constant of BL theory appropriately 
with a Lorentzian bi-invariant metric.
Then the Chern-Simons terms of BL theory become BF term and  
the kinetic terms for matter fields contain B-dependent terms besides 
other derivative terms.   
The simplest mass deformation to the BL Lagrangian 
is to add the single fermion mass term with modified supersymmetry 
transformations and other terms in the Lagrangian due to this
deformation term \cite{GSP,HLL}.
The aim of the first part 
in this section is to introduce the several mass terms 
for the fermion in the original BL Lagrangian. 
This procedure should preserve the exact ${\cal N}=2$
supersymmetry. We 
determine what is the correct expression for the bosonic
mass terms in the modified Lagrangian 
we should add to the original BL Lagrangian. 

Let us consider the deformed BL theory by adding 
four mass parameters $m_1, m_2, m_3$ and $m_4$ to the BL theory Lagrangian,
compared to \cite{HLL} where there are three mass parameters
\footnote{This paragraph is based on the discussion with K. Hosomichi 
intensively.}. See also the relevant paper by \cite{GSP} 
on the mass deformation.
Then the fermionic mass terms \footnote{The self-dual and anti
  self-dual tensors that are invariant under the $SU(3)_I \times
  U(1)_Y$ in ${\cal N}=8$ gauged supergravity
are given by
$
  X^{+}_{ijkl} = +[ (\de^{1234}_{ijkl}+\de^{5678}_{ijkl})+
 (\de^{1256}_{ijkl}+ \de^{3478}_{ijkl})+(\de^{1278}_{ijkl}
 +\de^{3456}_{ijkl})] $ and $ 
       X^{-}_{ijkl} = -[(\de^{1357}_{ijkl}
-\de^{2468}_{ijkl})+(\de^{1368}_{ijkl}
     -\de^{2457}_{ijkl})+(\de^{1458}_{ijkl} -\de^{2367}_{ijkl})-
   (\de^{1467}_{ijkl}-\de^{2358}_{ijkl})]$. The choice of \cite{HLL}
   for the mass parameters 
corresponds to the self-dual tensor for the indices $1234, 1256$, and
$1278$
while the choice of this paper
for the mass parameters  corresponds to the anti self-dual tensor 
for the indices $1357, 1368, 1458$ and $1467$ if we shift all the
   indices by adding $2$. For example, the indices $3689$ in
   (\ref{fermionic})
play the role of  1467 in above anti self-dual tensor.
\label{foot}  } 
from \cite{BL0711} are given by 
\bea
{\cal L}_{f.m.} =-\frac{i}{2} h_{ab} \bar{\Psi}^a \left(  m_1 \Gamma^{3579}+
  m_2\Gamma^{35810}+
  m_3\Gamma^{36710}-
  m_4\Gamma^{3689} \right) \Psi^b.
\label{fermionic}
\eea
Here the indices $a,b, \cdots$ run over the adjoint of the Lie algebra
for BL theory(and those indices run over the adjoint plus 
$+, -$  for BF Lorentzian model).
Then the corresponding  
fermionic supersymmetric transformation gets modified 
by
\bea
\delta_m \Psi^a = \left(  m_1 \Gamma^{3579}+
  m_2\Gamma^{35810}+
  m_3\Gamma^{36710}-
  m_4\Gamma^{3689} \right) X_I^a \Gamma_I \epsilon.
\label{mod} 
\eea
We impose three constraints on the $\epsilon$ parameter that satisfies the 
$\frac{1}{4}$ BPS condition(the number of supersymmetries is four)
$
\Gamma^{5678}\epsilon=\Gamma^{56910}\epsilon=
\Gamma^{78910}\epsilon=-\epsilon$ \footnote{These indices $5678,
  56910$ and $78910$ can be interpreted as $3456, 3478$ and $5678$ in
  $X^{+}_{ijkl}$ of footnote \ref{foot} respectively.}.
Now we introduce the bosonic mass term which preserves ${\cal N}=2$ 
supersymmetry and determine $(m^2)_{IJ}$:
\bea
{\cal L}_{b.m.} = -\frac{1}{2} h_{ab} X_I^a  (m^2)_{IJ} X_J^b.
\label{bosonic}
\eea
Using the supersymmetry variation for $X_I^a$,
$\delta X_I^a = i \bar{\epsilon} \Gamma_I \Psi^a$, and
the supersymmetry variation for $\Psi^a$ by the equation (\ref{mod}),
the variation for  the bosonic mass term (\ref{bosonic}) plus the
fermionic mass term (\ref{fermionic}) leads to
\bea
\delta {\cal L} =  i h_{ab} X_I^a  (m^2)_{IJ} \bar{\Psi}^b \Gamma_J \epsilon  
- i h_{ab} \bar{\Psi}^a  \left(  m_1 \Gamma^{3579}+
  m_2\Gamma^{35810}+
  m_3\Gamma^{36710}-
  m_4\Gamma^{3689} \right)^2
X_I^b \Gamma_I \epsilon. 
\label{van}
\eea
In order to vanish this, the bosonic mass term $(m^2)_{IJ} \Gamma_J$, by
computing the mass term for second term of (\ref{van}) explicitly \footnote{The
  relevant terms 
  become $m_1^2 +m_2^2 +m_3^2 +m_4^2 -2(m_1 m_4 +m_2 m_3)
  \Gamma^{5678}+
2(m_1 m_2 + m_3 m_4) \Gamma^{78910} + 2(m_1 m_3 + m_2 m_4)
\Gamma^{56910}$ explicitly.},
should take the form
\bea
&& (m_1-m_2-m_3+m_4)^2(\Gamma_3+\Gamma_4)
+(m_1-m_2+m_3-m_4)^2(\Gamma_5+\Gamma_6)
\nonu \\
 && +(m_1+m_2-m_3-m_4)^2(\Gamma_7+\Gamma_8)
+(m_1+m_2+m_3+m_4)^2(\Gamma_9+\Gamma_{10}).
\label{massdia}
\eea
In particular, when all the mass parameters are equal  
$
m_1 = m_2 =m_3 =m_4 \equiv m$, then
the diagonal bosonic mass term in (\ref{massdia}) has nonzero
components only for $99$  and $1010$ and other components($33, 44, 55,
66, 77$ and $88$) are vanishing \footnote{This resembles the structure of
$A_1^{IJ}$ tensor of $AdS_4$ supergravity where the $A_1^{IJ}$ tensor has two
distinct eigenvalues with degeneracies 6 and 2 respectively. The
degeneracy 2 is quite related to the ${\cal N}=2$ supersymmetry. For
the maximal supersymmetric case \cite{HLL}, all the diagonal 
mass matrix elements are equal and nonzero and this reflects the fact
that   the $A_1^{IJ}$ tensor has eight equal eigenvalues with degeneracies 8.     }: 
\bea
(m^2)_{IJ} = \diag(0,0,0,0,0,0,16m^2,16m^2).
\label{massmass}
\eea
Of course, the quartic terms for $X_I^a$ to the Lagrangian 
for our mass deformation can be fixed similarly, as in \cite{HLL}.
%Moreover, by adding the ghost sector \cite{GRVV} to our mass-deformed Lagrangian,
%the negative norm states can be removed. 
Let us introduce the four complex ${\cal N}=2$ superfields
as follows:
\bea
\Phi_1  & = &  X_3 + i X_4 + \cdots, 
\qquad \Phi_2 =X_5 + i X_6 + \cdots, 
\nonu \\
\Phi_3  & = &  X_7 + i X_8 + \cdots,  
\qquad \Phi_4 =X_9 + i X_{10} + \cdots
\label{n2}
\eea
where we do not include the ${\cal N}=2$ fermionic fields.
Or one can introduce these chiral superfields with 
an explicit $SU(4)_I$ fundamental representation as follows:
\bea
\Phi_A,  
%& = &  X_{A+2} + i X_{A+4} + \cdots 
\qquad A=(a,4), \qquad  a=1,2,3. 
\nonu
\eea
Recall from section 2 
that the ${\bf 8}$ of $SO(8)$ is relabelled by 
$A$ and $\bar{A}$ where $A=a, 4$ and $\bar{A}=\bar{a}, \bar{4}$.
Then the subset 
$\Phi_a$ where $a=1,2,3$ constitute a ${\bf 3}$ representation
of $SU(3)$ inside $SU(4)$.
The potential in the BL theory \cite{BL0711} is given by
\bea
\frac{1}{3\kappa^2} h_{ab} f_{cde}^{\;\;\;\;a} 
X_I^c X_J^d X_K^e f_{fgh}^{\;\;\;\;b} X_I^f X_J^g X_K^h
\nonu
\eea 
where $\kappa$ is a Chern-Simons coefficient.
In terms of ${\cal N}=2$ superfields, this contains the following expressions
\bea
\frac{2}{\kappa^2} h_{ab} f_{cde}^{\;\;\;\;a} f_{fgh}^{\;\;\;\;b}
\left[\Phi_1^c \Phi_2^d \Phi_3^e  \bar{\Phi}_1^f
\bar{\Phi}_2^g 
\bar{\Phi}_3^h + \mbox{three other terms} 
%\Phi_1^c \Phi_2^d \Phi_4^e  \bar{\Phi}_1^f
%\bar{\Phi}_2^g 
%\bar{\Phi}_4^h + \Phi_1^c \Phi_3^d \Phi_4^e  \bar{\Phi}_1^f
%\bar{\Phi}_3^g 
%\bar{\Phi}_4^h + \Phi_2^c \Phi_3^d \Phi_4^e  \bar{\Phi}_2^f
%\bar{\Phi}_3^g 
%\bar{\Phi}_4^h 
\right]
\nonu
\eea
by using the relation (\ref{n2}) 
between the component fields and superfields and a fundamental identity
is used.
This provides the superpotential:
$
\frac{\sqrt{2}}{\kappa} f_{abcd} f^{ABCD}
\Tr \Phi_A^a \Phi_B^b \Phi_C^c \Phi_D^d$.
Then this superpotential possesses $SU(4)_I$ global symmetry
%Note that the structure constant $f_{abcd}$ is nonzero only 
%if one of the four
%indices corresponds to $-$ index \cite{GMR,BRTV,HIM} 
%when we discuss about BF Lorentzian model 
\footnote{Note
that in \cite{BKKS} appeared in the same daily distribution of the arXiv, 
the ${\cal N}=2$ superspace formalism for BL
theory with gauge group $SU(2) \times SU(2)$ was found and the
superpotential has $SU(4)_I \times U(1)_R$ global symmetry. When the
normalization constants in the ${\cal N}=2$ superspace Lagrangian
hold some relation, the $R$-symmetry is enhanced to $SO(8)$
and further requirement on these constants allows this  ${\cal N}=2$ 
superspace Lagrangian to reduce to the one in component Lagrangian.
For BF Lorentzian model, it is not known yet how to write down the 
Lagrangian in ${\cal N}=2$ superspace formalism.
So it is not clear at this moment how one can proceed further on the
direction of BF Lorentzian model.
Furthermore, the mass deformed BL theory with two M2-branes 
is equivalent to the mass deformed $U(2) \times U(2)$
Chern-Simons gauge theory of \cite{ABJM} with level $k=1$ or $k=2$.}.

In ${\cal N}=2$ language, the superpotential consisting of the mass
term (\ref{massmass}) 
and quartic term, where we redefine $\Phi_4$ by diagonalizing the
mass matrix and introducing the new bosonic variables $X_9^a$ and
$X_{10}^a$,  is given by
\bea
W =  \frac{1}{2} M  h_{ab} \Tr \Phi_4^a \Phi_4^b + 
\frac{\sqrt{2}}{\kappa} f_{abcd} f^{ABCD}
\Tr \Phi_A^a \Phi_B^b \Phi_C^c \Phi_D^d.
\label{superW}
\eea
The global symmetry $SU(4)_I$ of $SO(8)$ is broken to $SU(3)_I$.
The second term is the superpotential required by 
${\cal N}=8$ supersymmetry as we mentioned above and 
the first term breaks ${\cal N}=8$ down to ${\cal N}=2$.
The theory has matter multiplet in three flavors $\Phi_1, \Phi_2$ and
$\Phi_3$
transforming in the adjoint. The $SO(8)_R$ symmetry of the 
${\cal N}=8$ gauge theory is broken to $SU(3)_I \times U(1)_R$
where the former is a flavor symmetry under which the matter multiplet
forms a triplet and the latter is the R-symmetry of the ${\cal N}=2$ theory.
Therefore, we turn on the mass perturbation in the UV and flow to the
IR.
This maps to turning on certain scalar fields in the $AdS_4$
supergravity where the scalars approach to zero in the UV($r
\rightarrow \infty$) and develop a nontrivial profile as a function of
$r$
becoming more significantly different from zero as one goes to the
IR($r \rightarrow -\infty$).
We can integrate out the massive scalar $\Phi_4$ with adjoint index 
at a low enough scale
and this results in the 6-th order superpotential $\Tr 
(f_{abc} f^{ABCD} \Phi_A^a 
\Phi_B^b \Phi_C^c)^2$.
% where one of the three
%indices has $-$ index \cite{GMR,BRTV,HIM} for BF Lorentzian model.  

The scale dimensions of four chiral superfields $\Phi_i(i=1,2,3,4)$ 
are $\Delta_i = \frac{1}{2}$ at the UV. 
This is because the sum of $\Delta_i$ is equal to the canonical dimension
of the superpotential which is $3-1=2$ \cite{Strassler98}.
By symmetry, one arrives at $\Delta_i=\frac{1}{2}$. 
The beta function from the mass term of $\Phi_4$ in (\ref{superW})
leads to $\beta_M = M (2\Delta_4 -2)$ \cite{JLP01}.
Or one can compute the anomalous mass dimension $\gamma_i$ explicitly as follows 
\cite{Strassler98}:
\bea
\beta_{1,1,1,1} & \sim & 4 \times (3-2) -2 \times (3-1) + 
\gamma_1 +\gamma_2
+\gamma_3 + \gamma_4 =  \gamma_1 +\gamma_2
+\gamma_3 + \gamma_4, \nonu \\
\beta_{0,0,0,2} & \sim & 2 \times (3-2) -2 \times (3-1) + 2
\gamma_4= -2 +2\gamma_4. 
\label{beta}
\eea 
The ${\cal N}=2$ supersymmetric gauge theory in three dimensions has a
holomorphic
superpotential and non-perturbative renormalizations of the
superpotential are restricted by holomorphy.
The form of (\ref{beta}) is a consequence of the non-renormalization
theorem for superpotential in ${\cal N}=2$ supersymmetry in three dimensions.
Then the vanishing of these (\ref{beta}) leads to $\gamma_1=\gamma_2=
\gamma_3=-\frac{1}{3}$ and 
$\gamma_4=1$.
This imposes one relation between $M$ and $\kappa$ suggesting that 
the theory has a fixed line of couplings.
Furthermore, the conformal dimension for $\Phi_4$ is given by
$\Delta_4=\frac{1}{2}(1+\gamma_4)=1$.
This comes from the relation $M(-2 + 2\Delta_4) = \frac{M}{2}(-2 +
2\gamma_4)$ using the equation (1) of \cite{Strassler98}.
Similarly, $\Delta_1=\Delta_2=\Delta_3=\frac{1}{3}$.
In other words, the IR values of scaling dimensions are $\Delta_4=1$ and 
$\Delta_i = \frac{1}{3}(i=1,2,3)$.
Then the $U(1)_R$ symmetry acts on $\Phi_1, \Phi_2, \Phi_3$ and
$\Phi_4$
with charges $(\frac{1}{3}, \frac{1}{3}, \frac{1}{3}, 1)$ which are
correctly related to the above anomalous dimensions. 
So both terms in the superpotential (\ref{superW}) have R charge 2, as
they must.
If we allow the mass $M$ to rotate by a phase then we have 
a further $U(1)$ symmetry under which 
$\Phi_i(i=1,2,3)$ has charge $\frac{1}{3}$ and $\Phi_4$ has charge
zero while the mass $M$ has charge 1.

In next section, the gauge invariant composites  
in the superconformal field theory at
the IR are mapped to the corresponding supergravity bulk fields.

%%%%%%%%%%%%%%%%%%%%%%%%%%%%%%%%%%%%%%%%%%%%%%%%%%%%%%%%%%%%%%%%
%%%%%%%%%%%%%%%%%%%%%%%%%%%%%%%%%%%%%%%%%%%%%%%%%%%%%%%%%%%%%%%%
\section{The $OSp(2|4)$ spectrum and operator map between bulk and
  boundary theories }
%%%%%%%%%%%%%%%%%%%%%%%%%%%%%%%%%%%%%%%%%%%%%%%%%%%%%%%%%%%%%%%%
%%%%%%%%%%%%%%%%%%%%%%%%%%%%%%%%%%%%%%%%%%%%%%%%%%%%%%%%%%%%%%%%

A further detailed correspondence between fields of
$AdS_4$ supergravity in four dimensions and composite operators of the 
IR field theory in three dimensions is described in this section.

The even subalgebra of the superalgebra $OSp(2|4)$
is a direct sum of subalgebras
where $Sp(4,R)\simeq SO(3,2)$ is the isometry algebra of $AdS_4$
and the compact subalgebra $SO(2)$ generates $U(1)_R$ symmetry
\cite{Merlatti}.
The maximally compact subalgebra is then 
$SO(2)_E \times SO(3)_S \times SO(2)_Y$
where the generator of $SO(2)_E$ is the hamiltonian of the system
and its eigenvalues $E$ are the energy levels of states for the
system,
the group $SO(3)_S$ is the roatation group and its representation $s$ 
describes the spin states of the system, and 
the eigenvalue $y$ of the generator of $SO(2)_Y$ is the hypercharge of
the state.

A supermultiplet, a unitary irreducible representations(UIR) of the
superalgebra
$OSp(2|4)$, consists of a finite number of UIR of the even subalgebra
and a particle state is characterized by a spin $s$, a mass $m$ and a 
hypercharge $y$. The relations between the mass and energy are given
in \cite{CFN} sometime ago.

Let us classify the supergravity multiplet which is invariant under 
$SU(3)_I \times U(1)_Y$ and describe them in the three dimensional 
boundary theory.

%%%%%%%%%%%%%%%%%%%%%%%%%%%%%%%%%%%%%%%%%
$\bullet$ Long massive vector multiplet
%%%%%%%%%%%%%%%%%%%%%%%%%%%%%%%%%%%%%%%%%

The conformal dimension $\Delta$, which is irrational and unprotected, 
is $\Delta=E_0$ and  
the $U(1)_R$ charge is $0$.
The $U(1)_R$ charge \footnote{The assignment of this $U(1)_R$ charge
  is different from the 
one given in \cite{AOT} where the $SO(8)$ branching rule is the same
as the present case because both theories have the same number of
supersymmetries.   
} is related to a hypercharge by
\bea
 R = y.
\label{ry}
\eea 
The $K(x, \theta^{+}, \theta^{-})$ is a general ``unconstrained''
scalar superfield in the boundary theory.
Since the Kahler potential evolves in the RG flow, 
the scalar field that measures 
the approach of the trajectory to the ${\cal N}=2$ point
sits in the supergravity 
multiplet dual to $K(x, \theta^{+}, \theta^{-})$, as in $AdS_5$
supergravity \cite{FGPW}.
This scalar field has a dimension $ \frac{1}{2}(5+\sqrt{17})$ in the IR.
We'll come back this issue at the end of this section.
The corresponding $OSp(2|4)$ representations and corresponding
${\cal N}=2$ superfield in three dimensions are listed in Table 2.
The relation between $\Delta$ and the mass for various fields can be
found in \cite{CFN}. For spin $0$ and $1$, their relations are 
given by $\Delta_{\pm} = 
\frac{3 \pm \sqrt{1+\frac{m^2}{4}}}{2}$ where we have to choose the
correct root among two cases as in \cite{Ahn99} while for spin $\frac{1}{2}$,
the explicit form is given by $\Delta =\frac{6 + |m|}{4}$. 
Using these relations, one can read off the mass for each state.

%%%%%%%%%%%%%%%%%%%%%%%%%%%%%%%%%%%%%%%%%%%%%%%%%%%%%%%%%%%%%%%%%%%%%%
%table 1%%%%%%%%%%%%%%%%%%%%%%%%%%%%%%%%%%%%%%%%%%%%%%%%%%%%%%%%%%%%%%%%%%%%%
\begin{table} 
\begin{center}
\begin{tabular}{|c|c|c|c|c|} \hline
Boundary Operator(BO) & Energy & Spin $0$  & Spin $\frac{1}{2}$ & Spin $1$
 \\ \hline
$K = \left( \Phi_1 \bar{\Phi}_1 +\Phi_2 \bar{\Phi}_2 +\Phi_3 \bar{\Phi}_3
\right)^{\frac{3}{2}}$ & $E_0=\frac{1}{2}(1+\sqrt{17})$ & ${\bf 1}_0$ &  &   \\
& $E_0+\frac{1}{2}=\frac{1}{2}(2+\sqrt{17})$ & & 
${\bf 1}_{\frac{1}{2}} \oplus {\bf
 1}_{-\frac{1}{2}}$  &
   \\
& $E_0+1=\frac{1}{2}(3+\sqrt{17})$ &  ${\bf 1}_1 \oplus {\bf 1}_0
 \oplus {\bf 1}_{-1} $ &  
& ${\bf 1}_0$  \\
& $E_0+\frac{3}{2}=
\frac{1}{2}(4+\sqrt{17})$ &  & ${\bf 1}_{\frac{1}{2}} 
\oplus {\bf 1}_{-\frac{1}{2}}$ &   \\
& $E_0+2 = \frac{1}{2}(5+\sqrt{17})$ & ${\bf 1}_0$ &  &   \\
\hline 
\end{tabular} 
\end{center}
\caption{\sl 
The $OSp(2|4)$ representations(energy, spin, hypercharge) 
and $SU(3)_I$ representations
in the supergravity mass spectrum for
long massive vector multiplet(corresponding to Table 1 of \cite{NW}) at
the ${\cal N}=2$ critical point and the corresponding ${\cal N}=2$
superfield in the boundary gauge theory.}
%\label{tableso4}
\end{table} 
%%%%%%%%%%%%%%%%%%%%%%%%%%%%%%%%%%%%%%%%%%%%%%%%%%%%%%%%%%%%%%%%%%%%%%%%%%%%
%%%%%%%%%%%%%%%%%%%%%%%%%%%%%%%%%%%%%%%%%%%%%%%%%%%%%%%%%%%%%%%%%%%%%%%%%%%%

%%%%%%%%%%%%%%%%%%%%%%%%%%%%%%%%%%%%%%%
$\bullet$ Short massive hypermultiplet
%%%%%%%%%%%%%%%%%%%%%%%%%%%%%%%%%%%%%%%

The conformal dimension $\Delta$ is the $U(1)_R$ charge for the lowest
component which can be written as $\Delta =\frac{E_0}{2}=|R|$.
The $AdS_4$ supergravity multiplet corresponds to the 
chiral scalar superfield $\Phi_c(x, \theta^{+})$ that 
satisfies $D^{+}_{\alpha} \Phi_c(x, \theta^{+}) = 0$ making the
multiplet short \cite{FFGRTZZ}
\footnote{The conformal dimension $\Delta$ and $U(1)_R$ charge for 
$\theta^{+}_{\alpha}$ are $\frac{1}{2}$ and $\frac{1}{2}$ while 
the conformal dimension $\Delta$ and $U(1)_R$ charge for 
$\theta^{-}_{\alpha}$ are $\frac{1}{2}$ and $-\frac{1}{2}$. }.
That is, in the $\theta^{+}$ expansion, there are three component
fields in the bulk. For the anti-chiral scalar superfield, one can see
the similar structure. 
Since the massive field $\Phi_4$ is integrated out in the flow, the IR
theory
contains the massless chiral superfields $\Phi_1, \Phi_2, \Phi_3$
with $\Delta =\frac{1}{3}$ and $U(1)_R$ charge $\frac{1}{3}$ from the
discussion of section 3 with (\ref{ry}). 
Then the bilinear of these chiral superfields by symmetrizing the two $SU(3)_I$
indices provides a symmetric representation of $SU(3)_I$, ${\bf 6 }$,
corresponding to $\Tr \Phi_{(i} \Phi_{j)}$
and its conjugate representation ${\bf \bar{6}}$, corresponding to $\Tr
\bar{\Phi}_{(i} \bar{\Phi}_{j)}$.
Using the relations between the dimension and mass for spin $0$ and 
$\frac{1}{2}$, one can also read off the mass for each state.
The corresponding $OSp(2|4)$ representations and corresponding
superfield are listed in Table 3.

%%%%%%%%%%%%%%%%%%%%%%%%%%%%%%%%%%%%%%%%%%%%%%%%%%%%%%%%%%%%%%%%%%%%%%%%%%
%table 2%%%%%%%%%%%%%%%%%%%%%%%%%%%%%%%%%%%%%%%%%%%%%%%%%%%%%%%%%%%%%%%%%%%%%%%%
\begin{table} 
\begin{center}
\begin{tabular}{|c|c|c|c|} \hline
Boundary Operator & Energy & Spin $0$  & Spin $\frac{1}{2}$ 
 \\ \hline
$\Tr \Phi_{(i} \Phi_{j)}$& 
$E_0=\frac{4}{3}$ & ${\bf 6}_{-\frac{2}{3}} 
\oplus {\bf \bar{6}}_{\frac{2}{3}}$ &     \\
complex & $E_0+ \frac{1}{2}= \frac{11}{6}$ & & 
${\bf 6}_{-\frac{1}{6}} \oplus {\bf
 \bar{6}}_{\frac{1}{6}}$     \\
& $E_0+1=
\frac{7}{3}$ &  ${\bf 6}_\frac{1}{3} \oplus 
{\bf \bar{6}}_{-\frac{1}{3}} $ &   \\
\hline 
\end{tabular} 
\end{center}
\caption{\sl 
The $OSp(2|4)$ representations(energy, spin, hypercharge) 
and $SU(3)_I$ representations in the supergravity mass spectrum for
short massive hypermultiplet(corresponding to Table 2 of \cite{NW}) at
the ${\cal N}=2$ critical point and the corresponding ${\cal N}=2$
superfield in the boundary gauge theory where $E_0=2|y|=2|R|$.}
%\label{tableso4}
\end{table} 
%%%%%%%%%%%%%%%%%%%%%%%%%%%%%%%%%%%%%%%%%%%%%%%%%%%%%%%%%%%%%%%%%%%%%%%%%%%%%
%%%%%%%%%%%%%%%%%%%%%%%%%%%%%%%%%%%%%%%%%%%%%%%%%%%%%%%%%%%%%%%%%%%%%%%%%%%%%

%%%%%%%%%%%%%%%%%%%%%%%%%%%%%%%%%%%%%%%%%%%
$\bullet$ Short massive gravitino multiplet
%%%%%%%%%%%%%%%%%%%%%%%%%%%%%%%%%%%%%%%%%%%

The conformal dimension $\Delta$ is the twice of $U(1)_R$ charge plus
$\frac{3}{2}$ for the lowest component, $\Delta = E_0=2|R| +\frac{3}{2}$.
This  corresponds to 
spinorial superfield $\Phi_{\alpha}(x, \theta^{+})$ that 
satisfies $D^{+ \alpha} \Phi_{\alpha}(x, \theta^{+}) = 0$ \cite{FFGT}. 
Of course, this constraint makes the multiplet short. 
In the $\theta^{\pm}$ expansion, the component
fields in the bulk are located with appropriate quantum numbers. 
The massless chiral superfields $\Phi_1, \Phi_2, \Phi_3$
have $\Delta =\frac{1}{3}$ and $U(1)_R$ charge $\frac{1}{3}$ as before. 
The gauge
superfield $W_{\alpha}$ has $\Delta =\frac{3}{2}$ 
and $U(1)_R$ charge $-\frac{1}{6}$ and its conjugate field has
opposite $U(1)_R$ charge $\frac{1}{6}$.  
Then one can identify $\Tr W_{\alpha} \Phi_j$ with ${\bf 3}$
and $\Tr \bar{W}_{\alpha} \bar{\Phi}_j$ with ${\bf \bar{3}}$.
The corresponding $OSp(2|4)$ representations and corresponding
superfield are listed in Table 4.
For spin $\frac{3}{2}$, the relation for the mass and dimension is
given by
$\Delta =\frac{6 + |m+4|}{4}$ and for spin $0, 1$ and $\frac{1}{2}$,
the previous relations hold.

%%%%%%%%%%%%%%%%%%%%%%%%%%%%%%%%%%%%%%%%%%%%%%%%%%%%%%%%%%%%%%
%table 3%%%%%%%%%%%%%%%%%%%%%%%%%%%%%%%%%%%%%%%%%%%%%%%%%%%%%%%%%%%%
\begin{table} 
\begin{center}
\begin{tabular}{|c|c|c|c|c|c|} \hline
B.O. & Energy & Spin $0$  & Spin $\frac{1}{2}$ & Spin $1$
 & Spin $\frac{3}{2}$ \\ \hline 
$\Tr W_{\alpha} \Phi_j$ 
& $E_0
=\frac{11}{6}$ &  & ${\bf 3}_{\frac{1}{6}} 
\oplus {\bf \bar{3}}_{-\frac{1}{6}}$ &  & \\
complex & $E_0 +\frac{1}{2} = \frac{7}{3}$ & 
${\bf 3}_{-\frac{1}{3}} \oplus {\bf
 \bar{3}}_{\frac{1}{3}}$  & & ${\bf 3}_{\frac{2}{3}} \oplus {\bf
 3}_{-\frac{1}{3}} \oplus {\bf \bar{3}}_{-\frac{2}{3}} \oplus
{\bf \bar{3}}_{\frac{1}{3}} $ &  \\
& $E_0+1 = \frac{17}{6}$ & & ${\bf 3}_{\frac{1}{6}} \oplus {\bf
 \bar{3}}_{-\frac{1}{6}} \oplus {\bf 3}_{-\frac{5}{6}} \oplus
{\bf \bar{3}}_{\frac{5}{6}} $ 
&   & ${\bf 3}_{\frac{1}{6}} \oplus {\bf \bar{3}}_{-\frac{1}{6}}$  \\
& $E_0+\frac{3}{2} = \frac{10}{3}$ &  &  & ${\bf 3}_{-\frac{1}{3}} 
\oplus {\bf \bar{3}}_{\frac{1}{3}}$ &  \\
\hline 
\end{tabular} 
\end{center}
\caption{\sl 
The $OSp(2|4)$ representations(energy, spin, hypercharge) 
and  $SU(3)_I$ representations in the supergravity mass spectrum for
short massive gravitino multiplet(corresponding to Table 3 of
\cite{NW}) at
the ${\cal N}=2$ critical point and the corresponding ${\cal N}=2$
superfield in the boundary gauge theory where $E_0= 2|y|+\frac{3}{2} =
2|R|+\frac{3}{2}$.}
%\label{tableso4}
\end{table} 
%%%%%%%%%%%%%%%%%%%%%%%%%%%%%%%%%%%%%%%%%%%%%%%%%%%%%%%%%%%%%%%%%%%%%%%%%%
%%%%%%%%%%%%%%%%%%%%%%%%%%%%%%%%%%%%%%%%%%%%%%%%%%%%%%%%%%%%%%%%%%%%%%%%%%

%%%%%%%%%%%%%%%%%%%%%%%%%%%%%%%%%%%%%%%%%%%%%%%%%%
$\bullet$ ${\cal N}=2$ massless graviton multiplet
%%%%%%%%%%%%%%%%%%%%%%%%%%%%%%%%%%%%%%%%%%%%%%%%%%

This can be identified with the  
stress energy tensor superfield $T^{\alpha \beta}(x, \theta^{+},\theta^{-})$
that satisfies the equations $D_{\alpha}^{+} T^{\alpha \beta}=0 
=D_{\alpha}^{-} T^{\alpha \beta}$ 
\cite{FFGRTZZ,Ahn02-3}. In components, 
the $\theta^{\pm}$ expansion of this superfield has
the stress energy tensor, the ${\cal N}=2$ supercurrents, and $U(1)_R$
symmetry current.
The conformal dimension $\Delta=2$ and the $U(1)_R$ charge is $0$.
This has protected dimension.
The corresponding $OSp(2|4)$ representations and corresponding
superfield are listed in Table 5.
 For spin $2$, we have the relation $\Delta_{\pm} = 
\frac{3 \pm \sqrt{9+\frac{m^2}{4}}}{2}$ and for massless case, this
leads to $\Delta_{+}=3$.

%%%%%%%%%%%%%%%%%%%%%%%%%%%%%%%%%%%%%%%%%%%%%%%%%%
$\bullet$ ${\cal N}=2$ massless vector multiplet
%%%%%%%%%%%%%%%%%%%%%%%%%%%%%%%%%%%%%%%%%%%%%%%%%%

This conserved vector current is given by 
a scalar superfield $J^A(x, \theta^{+}, \theta^{-})$ that 
satisfies $D^{+\alpha} D^{+}_{\alpha} J^A = 0=
D^{-\alpha} D^{-}_{\alpha} J^A$ \cite{FFGRTZZ}. 
This transforms in the
adjoint representation of $SU(3)_I$ flavor group. The boundary object
is given by 
$\Tr  \bar{\Phi} T^A \Phi$ where 
the flavor indices in $\Phi_i$ and $\bar{\Phi}_i$ are contracted
and the generator $T^A$ is   $N \times
N$ matrix  with $A=1, 2, \cdots, N^2-1$. 
The conformal dimension $\Delta=1$ and the $U(1)_R$ charge is $0$.
This has also protected dimension.
By taking a tensor product between ${\bf 3}$ and ${\bf \bar{3}}$, one
gets this octet ${\bf 8}$ of $SU(3)_I$ representation.
The corresponding $OSp(2|4)$ representations and corresponding
superfield are listed in Table 5 also.

%%%%%%%%%%%%%%%%%%%%%%%%%%%%%%%%%%%%%%%%%%%%%%%%%%%%%%%%%%%%%%%%
% table 4%%%%%%%%%%%%%%%%%%%%%%%%%%%%%%%%%%%%%%%%%%%%%%%%%%%%%%%%%%%%%
\begin{table} 
\begin{center}
\begin{tabular}{|c|c|c|c|c|c|c|} \hline
Boundary Operator & Energy & Spin $0$  & Spin $\frac{1}{2}$ & Spin $1$
 & Spin $\frac{3}{2}$ & Spin $2$ \\ \hline 
$\Tr  \bar{\Phi} T^A \Phi $ 
& $E_0
=1$ & ${\bf 8}_0 $  
&  &  &  & \\
& $E_0 +\frac{1}{2} = \frac{3}{2}$ &   & ${\bf 8}_{\frac{1}{2}} \oplus
 {\bf 8}_{-\frac{1}{2}}$ &  & & \\
& $E_0+1 = 2$ &  ${\bf 8}_0$ &  
& ${\bf 8}_0$  &  & \\
\hline
$T^{\alpha \beta}$ & $E_0=2$ & & & ${\bf 1}_0$ & & \\
 & $E_0+\frac{1}{2}=\frac{5}{2}$ 
& & & & ${\bf 1}_{\frac{1}{2}} \oplus {\bf 1}_{-\frac{1}{2}}$ & \\
& $E_0 +1 = 3$ & & & & & ${\bf 1}_0$ \\
\hline
\end{tabular} 
\end{center}
\caption{\sl 
The $OSp(2|4)$ representations(energy, spin, hypercharge) 
and  $SU(3)_I$ representations in the supergravity mass spectrum for
``ultra'' short multiplets at
the ${\cal N}=2$ critical point and the corresponding ${\cal N}=2$
superfields in the boundary gauge theory.}
%\label{tableso4}
\end{table} 
%%%%%%%%%%%%%%%%%%%%%%%%%%%%%%%%%%%%%%%%%%%%%%%%%%%%%%%%%%%%%%%%%%%%
%%%%%%%%%%%%%%%%%%%%%%%%%%%%%%%%%%%%%%%%%%%%%%%%%%%%%%%%%%%%%%%%%%%%

Let us describe the Kahler potential more detail we mentioned in 
the long vector multiplet.
The Kahler potential 
is found in \cite{JLP01}, by looking at the 11 dimensional flow
equation \cite{CPW},  as 
\bea
K = \frac{1}{4} \tau_{M2} L^2 e^A \left( \rho^2 + \frac{1}{\rho^6}
\right), \qquad \frac{d q}{d r} = \frac{2}{L \rho^2} q
\label{sol}
\eea
where $\rho \equiv e^{\frac{\lambda}{4\sqrt{2}}}$ and $\chi \equiv 
\frac{\lambda'}{\sqrt{2}}$.
The corresponding Kahler metric is given by \cite{JLP01}
\bea
d s^2 = \frac{1}{4q^2} \left( q \frac{d}{d q}\right)^2 K dq^2 +
\left( q \frac{d}{d q} \right) K d \hat{x}^I d \hat{x}^I +
\left( q^2 \frac{d^2}{ d q^2} \right) K (\hat{x}^I J_{IJ} d \hat{x}^J )^2
\label{metric}
\eea
where the coordinate $q$ is defined as
$
q   \equiv  w^1 \bar{w}^1 + w^2 \bar{w}^2 + w^3 \bar{w}^3$ and 
the three complex coordinates are given by
$w^1   =  \sqrt{q} \left( \hat{x}^1 + i \hat{x}^2 \right), 
w^2 = \sqrt{q} \left( \hat{x}^3 + i \hat{x}^4 \right)$ and
$w^3 = \sqrt{q} \left( \hat{x}^5 + i \hat{x}^6 \right)$ on ${\bf C}^3$
and the $\hat{x}$'s are coordinates on an ${\bf S}^5$ of unit radius. 
So we reparametrize ${\bf C}^3$ with coordinates $\hat{x}^1, \cdots, 
\hat{x}^6$ and $q$.
Here $J$ is an antisymmetric matrix with $J_{12} =J_{34}=J_{56}=1$.
The $d \hat{x}^I d \hat{x}^I$ is a metric on a round ${\bf S}^5$ and 
$(\hat{x}^I J_{IJ} d \hat{x}^J)^2$ is the $U(1)$ fiber in the
description of ${\bf S}^5$. 
Note that there is a relation $\frac{d K}{d r} = \tau_{M_2} 
L e^A$ \cite{JLP01}.
The moduli space is parametrized by the vacuum expectation values of
the three massless scalars $\Phi_1, \Phi_2$ and $\Phi_3$ denoted as
$w^1, w^2$ and $w^3$.  The $w^i(i=1, 2, 3)$ transform in the
fundamental representation ${\bf 3}$ of $SU(3)_I$ while their complex
conjugates $\bar{w}^i$ transform in the anti-fundamental representation 
${\bf \bar{3}}$.  

At the UV end of the flow which is just $AdS_4 \times {\bf S}^7$, 
$A(r) \sim \frac{2}{L} r$
from the solution (\ref{solution})
for $A(r)$ and $W=1$ from Table 1. Moreover,
the radial coordinate on moduli space $\sqrt{q} \sim
e^{\frac{r}{ L}} \sim e^{\frac{A(r)}{2}}
$ 
from (\ref{sol}) by
substituting $\rho =1$ from Table 1. Therefore, the
Kahler potential from (\ref{sol}) behaves as $K \sim e^{A(r)} \sim q $. 
This implies that  $K = \Phi_1 \bar{\Phi}_1 +\Phi_2 \bar{\Phi}_2+
\Phi_3 \bar{\Phi}_3 $ at the UV in the boundary theory.
Since the scaling dimensions for $\Phi_i(i=1,2,3)$ and its conjugate
fields
are $\frac{1}{2}$, the scaling dimension of $K$ is equal to $1$
which is correct because it should have scaling dimension $1$ 
``classically'' from $\int d^3 x \pa_{\varphi} \pa_{\bar{\varphi}} K
\pa_\mu \varphi \pa^\mu \bar{\varphi}$ where $\varphi$ are the
massless scalars with some scaling dimensions.

At the IR end of the flow, $A(r) \sim \frac{3^{\frac{3}{4}}}{L} r$
with $g \equiv \frac{\sqrt{2}}{L}$ from the solution (\ref{solution})
for $A(r)$ and $W= \frac{3^{\frac{3}{4}}}{2}$ from Table 1. Moreover,
$\sqrt{q} \sim
e^{\frac{3^{-\frac{1}{4}} r}{ L}} \sim e^{\frac{A(r)}{3}}
$ 
from (\ref{sol}) by
substituting $\rho =3^{\frac{1}{8}}$ from Table 1. Therefore, the
Kahler potential behaves as $K \sim e^{A(r)} \sim q^{\frac{3}{2}}$. 
Then $K$ becomes $K =(\Phi_1 \bar{\Phi}_1 +\Phi_2 \bar{\Phi}_2+
\Phi_3 \bar{\Phi}_3 )^{\frac{3}{2}}$ in the boundary theory.
Obviously, from the tensor product between ${\bf 3}$ and ${\bf \bar{3}}$
of $SU(3)_I$ representation, one gets a singlet ${\bf 1}_0$ with
$U(1)_R$ charge $0$. Note that $\Phi_i(i=1,2,3)$ has $U(1)_R$ charge 
$\frac{1}{3}$ while $\bar{\Phi}_i(i=1,2,3)$ has $U(1)_R$ charge 
$-\frac{1}{3}$.
Since the scaling dimensions for 
 $\Phi_i(i=1,2,3)$ and its conjugate
fields
are $\frac{1}{3}$, the scaling dimension of $K$ is $1$ which is
consistent with ``classical'' value as before.
The corresponding Kahler metric (\ref{metric}) 
provides the Kahler term in the action.
For the superfield $K(x, \theta^{+}, \theta^{-})$, the 
action looks like $\int d^3 x d^2 \theta^{+} d^2 \theta^{-} 
K(x, \theta^{+}, \theta^{-})$. The component content of this action 
can be worked out straightforwardly using the projection technique. 
This implies that the highest component field in $\theta^{\pm}$-expansion,
the last element in Table 2, has a conformal dimension  $\frac{1}{2} 
(5+\sqrt{17})$ in the IR as before \footnote{So far we have considered
the leading behavior of Kahler potential at the two end points of UV
and IR. This can be understood from the ``classical'' description of
scaling dimension above also. However, one can look at
next-to-leading order ``quantum'' 
corrections to this Kahler potential. The exact
expression for the Kahler potential along ``the whole flow'' 
is given by (\ref{sol}). One can
easily obtain the asymptotic behaviors of $A(r)$ and $\rho(r)$ 
around IR region. The former can be determined through the last
one of first
order differential equations (\ref{solution}) by expanding the
superpotential $W$ around $\rho =3^{\frac{1}{8}}$ and $\chi=\frac{1}{2} 
\cosh^{-1} 2$ while the latter can be obtained through the first
equation of (\ref{solution}) by expanding the right hand side of that
equation around IR fixed point values 
$\rho =3^{\frac{1}{8}}$ and $\chi=\frac{1}{2} 
\cosh^{-1} 2$. Then one expects that the irrational piece $3-\sqrt{17}$
from the mass spectrum found in \cite{AP} 
arises in the exponent of next-to-leading order $r$-dependent term in
$\rho$
and $\chi$. The coefficient appearing in the next-to-leading order of Kahler
potential is related to the mass of $\Phi_4$ via 
M2-brane probe analysis. 
One can approximate the Kahler potential by 
$K \sim (\Phi_1 \bar{\Phi}_1 +\Phi_2 \bar{\Phi}_2+
\Phi_3 \bar{\Phi}_3 )^{\frac{3}{2}}$ up to leading order at ``the IR fixed point''
but along the flow around the IR, in general,
the Kahler potential is given by (\ref{sol}). 
For the relevant work on $AdS_5 \times {\bf S}^5$
compactification with D3-branes, see the section 
2.5 of \cite{JLP01} for example.}. 

We have presented the gauge invariant 
combinations of the massless superfields 
of the gauge theory whose scaling dimensions and $SU(3)_I \times
U(1)_R$ quantum numbers exactly match the four short multiplets in
Tables $3,4, 5$
observed in the supergravity.  There exists one additional long
multiplet in Table $2$ which completes the picture. 

%%%%%%%%%%%%%%%%%%%%%%%%%%%%%%%%%%%%%%%%%%%%%%%%%%%%%%%%%%%%%%%%%%%%%%%%%%%%%%%
%%%%%%%%%%%%%%%%%%%%%%%%%%%%%%%%%%%%%%%%%%%%%%%%%%%%%%%%%%%%%%%%%%%%%%%%%%%%%%%%
\section{
Conclusions and outlook }
%%%%%%%%%%%%%%%%%%%%%%%%%%%%%%%%%%%%%%%%%%%%%%%%%%%%%%%%%%%%%%%%%%%%%%%%%%%%%%%%
%%%%%%%%%%%%%%%%%%%%%%%%%%%%%%%%%%%%%%%%%%%%%%%%%%%%%%%%%%%%%%%%%%%%%%%%%%%%%%%%

By studying the mass-deformed Bagger-Lambert theory, preserving 
$SU(3)_I \times U(1)_R$ symmetry, with the addition of mass term
for one of the four adjoint chiral superfields, 
one identifies an ${\cal N}=2$ supersymmetric membrane flow in three
dimensional deformed BL theory with 
the holographic ${\cal N}=2$ 
supersymmetric RG flow in four dimensions. 
Therefore, the ${\cal N}=8$ gauged supergravity critical point
is indeed the holographic dual of the mass-deformed ${\cal N}=8$ BL theory.
So far, we have focused on the particular mass deformation 
(\ref{fermionic}) preserving 
$SU(3)_I \times U(1)_R$ symmetry. It would be interesting 
to discover all the possible classification for the mass deformations
and see how they appear in the $AdS_4 \times {\bf S}^7$ background where
some of them are nonsupersymmetric and some of them are 
supersymmetric \cite{AT02-1}.  

\vspace{.7cm}

%%%%%%%%%%%%%%%%%%%%%%%%%%%%%%%%%%%%%%%%%%%%%%%%%%%%%%%%%%%%%%
%%%%%%%%%%%%%%%%%%%%%%%%%%%%%%%%%%%%%%%%%%%%%%%%%%%%%%%%%%%%%%%
\centerline{\bf Acknowledgments}
%%%%%%%%%%%%%%%%%%%%%%%%%%%%%%%%%%%%%%%%%%%%%%%%%%%%%%%%%%%%%%%
%%%%%%%%%%%%%%%%%%%%%%%%%%%%%%%%%%%%%%%%%%%%%%%%%%%%%%%%%%%%%%%

I would like to thank 
Kazuo Hosomichi and Sungjay Lee  
for discussions. 
This work was supported by grant No.
R01-2006-000-10965-0 from the Basic Research Program of the Korea
Science \& Engineering Foundation.  
I acknowledge D. Kutasov for warm hospitality of 
Particle Theory Group, Enrico Fermi Institute 
at University of Chicago.
%where this work was initiated.
%I would like to thank KIAS(Korea Institute for 
%Advanced Study) for hospitality.  
%where
%this work was undertaken. 


\begin{thebibliography}{99}

%\cite{Maldacena:1997re}
\bibitem{Maldacena}
  J.~M.~Maldacena,
%  ``The large N limit of superconformal field theories and supergravity,''
  Adv.\ Theor.\ Math.\ Phys.\  {\bf 2}, 231 (1998)
  [Int.\ J.\ Theor.\ Phys.\  {\bf 38}, 1113 (1999)]
  [arXiv:hep-th/9711200].
  %%CITATION = IJTPB,38,1113;%%

%\cite{Seiberg:1997ax}
\bibitem{Seiberg}
  N.~Seiberg,
%  ``Notes on theories with 16 supercharges,''
  Nucl.\ Phys.\ Proc.\ Suppl.\  {\bf 67}, 158 (1998)
  [arXiv:hep-th/9705117].
  %%CITATION = NUPHZ,67,158;%%

%\cite{Ahn:2000aq}
\bibitem{AP}
  C.~Ahn and J.~Paeng,
%  ``Three-dimensional SCFTs, supersymmetric domain wall and renormalization
%  group flow,''
  Nucl.\ Phys.\  B {\bf 595}, 119 (2001)
  [arXiv:hep-th/0008065].
  %%CITATION = NUPHA,B595,119;%%

%\cite{Ahn:2000mf}
\bibitem{AW}
  C.~Ahn and K.~Woo,
%  ``Supersymmetric domain wall and RG flow from 4-dimensional gauged N = 8
%  supergravity,''
  Nucl.\ Phys.\  B {\bf 599}, 83 (2001)
  [arXiv:hep-th/0011121].
  %%CITATION = NUPHA,B599,83;%%

%\cite{Corrado:2001nv}
\bibitem{CPW}
  R.~Corrado, K.~Pilch and N.~P.~Warner,
%  ``An N = 2 supersymmetric membrane flow,''
  Nucl.\ Phys.\  B {\bf 629}, 74 (2002)
  [arXiv:hep-th/0107220].
  %%CITATION = NUPHA,B629,74;%%

%\cite{Johnson:2001ze}
\bibitem{JLP01}
  C.~V.~Johnson, K.~J.~Lovis and D.~C.~Page,
%  ``The Kaehler structure of supersymmetric holographic RG flows,''
  JHEP {\bf 0110}, 014 (2001)
  [arXiv:hep-th/0107261].
  %%CITATION = JHEPA,0110,014;%%

%\cite{Bagger:2007jr}
\bibitem{BL0711}
  J.~Bagger and N.~Lambert,
%  ``Gauge Symmetry and Supersymmetry of Multiple M2-Branes,''
  Phys.\ Rev.\  D {\bf 77}, 065008 (2008)
  [arXiv:0711.0955 [hep-th]].
  %%CITATION = PHRVA,D77,065008;%%

%\cite{Bagger:2006sk}
\bibitem{BL06}
  J.~Bagger and N.~Lambert,
  %``Modeling multiple M2's,''
  Phys.\ Rev.\  D {\bf 75}, 045020 (2007)
  [arXiv:hep-th/0611108].
  %%CITATION = PHRVA,D75,045020;%%

%\cite{Bagger:2007vi}
\bibitem{BL0712}
  J.~Bagger and N.~Lambert,
%  ``Comments On Multiple M2-branes,''
  JHEP {\bf 0802}, 105 (2008)
  [arXiv:0712.3738 [hep-th]].
  %%CITATION = JHEPA,0802,105;%%

%\cite{Gustavsson:2007vu}
\bibitem{Gustavsson07}
  A.~Gustavsson,
%  ``Algebraic structures on parallel M2-branes,''
  arXiv:0709.1260 [hep-th].
  %%CITATION = ARXIV:0709.1260;%%

%\cite{Gustavsson:2008dy}
\bibitem{Gustavsson08}
  A.~Gustavsson,
%  ``Selfdual strings and loop space Nahm equations,''
  JHEP {\bf 0804}, 083 (2008)
  [arXiv:0802.3456 [hep-th]].
  %%CITATION = JHEPA,0804,083;%%

%\cite{Gomis:2008uv}
\bibitem{GMR}
  J.~Gomis, G.~Milanesi and J.~G.~Russo,
%  ``Bagger-Lambert Theory for General Lie Algebras,''
  arXiv:0805.1012 [hep-th].
  %%CITATION = ARXIV:0805.1012;%%

%\cite{Benvenuti:2008bt}
\bibitem{BRTV}
  S.~Benvenuti, D.~Rodriguez-Gomez, E.~Tonni and H.~Verlinde,
%  ``N=8 superconformal gauge theories and M2 branes,''
  arXiv:0805.1087 [hep-th].
  %%CITATION = ARXIV:0805.1087;%%

%\cite{Ho:2008ei}
\bibitem{HIM}
  P.~M.~Ho, Y.~Imamura and Y.~Matsuo,
%  ``M2 to D2 revisited,''
  arXiv:0805.1202 [hep-th].
  %%CITATION = ARXIV:0805.1202;%%

%%%%%%%%%%%%%%%%%%%%%%%%%%%%%%%%%%%%%%%%%%%%%%%%%%%%%%%%%%%%%%%%%%%%%%%%%%%%%%%%%
%%%%%%%%%%%%%%%%%%%%%%%%%%%%%%%%%%%%%%%%%%%%%%%%%%%%%%%%%%%%%%%%%%%%%%%%%%%%%%%%%

%\cite{Aharony:2008ug}
\bibitem{ABJM}
  O.~Aharony, O.~Bergman, D.~L.~Jafferis and J.~Maldacena,
  %``N=6 superconformal Chern-Simons-matter theories, M2-branes and their
  %gravity duals,''
  arXiv:0806.1218 [hep-th].
  %%CITATION = ARXIV:0806.1218;%%

%\cite{FigueroaO'Farrill:2008fz}
\bibitem{Figueroa}
  J.~Figueroa-O'Farrill,
  %``Lorentzian Lie n-algebras,''
  arXiv:0805.4760 [math.RT].
  %%CITATION = ARXIV:0805.4760;%%

%\cite{Gomis:2008be}
\bibitem{GRVV}
  J.~Gomis, D.~Rodriguez-Gomez, M.~Van Raamsdonk and H.~Verlinde,
  %``The Superconformal Gauge Theory on M2-Branes,''
  arXiv:0806.0738 [hep-th].
  %%CITATION = ARXIV:0806.0738;%%

%\cite{Passerini:2008qt}
\bibitem{Pass}
  F.~Passerini,
  %``M2-Brane Superalgebra from Bagger-Lambert Theory,''
  arXiv:0806.0363 [hep-th].
  %%CITATION = ARXIV:0806.0363;%%

%\cite{Park:2008qe}
\bibitem{Park:2008qe}
  J.~H.~Park and C.~Sochichiu,
  %``Single M5 to multiple M2: taking off the square root of Nambu-Goto
  %action,''
  arXiv:0806.0335 [hep-th].
  %%CITATION = ARXIV:0806.0335;%%

%\cite{Bandres:2008kj}
\bibitem{Bandres:2008kj}
  M.~A.~Bandres, A.~E.~Lipstein and J.~H.~Schwarz,
  %``Ghost-Free Superconformal Action for Multiple M2-Branes,''
  arXiv:0806.0054 [hep-th].
  %%CITATION = ARXIV:0806.0054;%%

%\cite{Gustavsson:2008bf}
\bibitem{Gustavsson:2008bf}
  A.~Gustavsson,
  %``One-loop corrections to Bagger-Lambert theory,''
  arXiv:0805.4443 [hep-th].
  %%CITATION = ARXIV:0805.4443;%%

%\cite{FigueroaO'Farrill:2008zm}
\bibitem{FigueroaO'Farrill:2008zm}
  J.~Figueroa-O'Farrill, P.~de Medeiros and E.~Mendez-Escobar,
  %``Lorentzian Lie 3-algebras and their Bagger-Lambert moduli space,''
  arXiv:0805.4363 [hep-th].
  %%CITATION = ARXIV:0805.4363;%%

%\cite{Lin:2008qp}
\bibitem{Lin:2008qp}
  H.~Lin,
  %``Kac-Moody Extensions of 3-Algebras and M2-branes,''
  arXiv:0805.4003 [hep-th].
  %%CITATION = ARXIV:0805.4003;%%

%\cite{Banerjee:2008pd}
\bibitem{Banerjee:2008pd}
  S.~Banerjee and A.~Sen,
%  ``Interpreting the M2-brane Action,''
  arXiv:0805.3930 [hep-th].
  %%CITATION = ARXIV:0805.3930;%%

%\cite{Hosomichi:2008jd}
\bibitem{Hosomichi:2008jd}
  K.~Hosomichi, K.~M.~Lee, S.~Lee, S.~Lee and J.~Park,
%  ``N=4 Superconformal Chern-Simons Theories with Hyper and Twisted Hyper
%  Multiplets,''
  arXiv:0805.3662 [hep-th].
  %%CITATION = ARXIV:0805.3662;%%

%\cite{Li:2008ez}
\bibitem{Li:2008ez}
  M.~Li and T.~Wang,
%  ``M2-branes Coupled to Antisymmetric Fluxes,''
  arXiv:0805.3427 [hep-th].
  %%CITATION = ARXIV:0805.3427;%%

%\cite{Jeon:2008bx}
\bibitem{Jeon:2008bx}
  I.~Jeon, J.~Kim, N.~Kim, S.~W.~Kim and J.~H.~Park,
%  ``Classification of the BPS states in Bagger-Lambert Theory,''
  arXiv:0805.3236 [hep-th].
  %%CITATION = ARXIV:0805.3236;%%

%\cite{Song:2008bi}
\bibitem{Song:2008bi}
  Y.~Song,
%  ``Mass Deformation of the Multiple M2 Branes Theory,''
  arXiv:0805.3193 [hep-th].
  %%CITATION = ARXIV:0805.3193;%%

%\cite{Krishnan:2008zm}
\bibitem{Krishnan:2008zm}
  C.~Krishnan and C.~Maccaferri,
%  ``Membranes on Calibrations,''
  arXiv:0805.3125 [hep-th].
  %%CITATION = ARXIV:0805.3125;%%

%\cite{Ho:2008ve}
\bibitem{Ho:2008ve}
  P.~M.~Ho, Y.~Imamura, Y.~Matsuo and S.~Shiba,
%  ``M5-brane in three-form flux and multiple M2-branes,''
  arXiv:0805.2898 [hep-th].
  %%CITATION = ARXIV:0805.2898;%%

%\cite{Fuji:2008yj}
\bibitem{Fuji:2008yj}
  H.~Fuji, S.~Terashima and M.~Yamazaki,
%  ``A New N=4 Membrane Action via Orbifold,''
  arXiv:0805.1997 [hep-th].
  %%CITATION = ARXIV:0805.1997;%%

%\cite{Honma:2008un}
\bibitem{Honma:2008un}
  Y.~Honma, S.~Iso, Y.~Sumitomo and S.~Zhang,
%  ``Janus field theories from multiple M2 branes,''
  arXiv:0805.1895 [hep-th].
  %%CITATION = ARXIV:0805.1895;%%

%\cite{Morozov:2008rc}
\bibitem{Morozov:2008rc}
  A.~Morozov,
%  ``From Simplified BLG Action to the First-Quantized M-Theory,''
  arXiv:0805.1703 [hep-th].
  %%CITATION = ARXIV:0805.1703;%%

%\cite{Ho:2008nn}
\bibitem{Ho:2008nn}
  P.~M.~Ho and Y.~Matsuo,
  %``M5 from M2,''
  arXiv:0804.3629 [hep-th].
  %%CITATION = ARXIV:0804.3629;%%

%\cite{Papadopoulos:2008gh}
\bibitem{Papadopoulos:2008gh}
  G.~Papadopoulos,
%  ``On the structure of k-Lie algebras,''
  arXiv:0804.3567 [hep-th].
  %%CITATION = ARXIV:0804.3567;%%

%\cite{Gauntlett:2008uf}
\bibitem{Gauntlett:2008uf}
  J.~P.~Gauntlett and J.~B.~Gutowski,
%  ``Constraining Maximally Supersymmetric Membrane Actions,''
  arXiv:0804.3078 [hep-th].
  %%CITATION = ARXIV:0804.3078;%%

%\cite{Papadopoulos:2008sk}
\bibitem{Papadopoulos:2008sk}
  G.~Papadopoulos,
%  ``M2-branes, 3-Lie Algebras and Plucker relations,''
  JHEP {\bf 0805}, 054 (2008)
  [arXiv:0804.2662 [hep-th]].
  %%CITATION = JHEPA,0805,054;%%

%\cite{Bergshoeff:2008cz}
\bibitem{Bergshoeff:2008cz}
  E.~A.~Bergshoeff, M.~de Roo and O.~Hohm,
%  ``Multiple M2-branes and the Embedding Tensor,''
  arXiv:0804.2201 [hep-th].
  %%CITATION = ARXIV:0804.2201;%%

%\cite{Ho:2008bn}
\bibitem{Ho:2008bn}
  P.~M.~Ho, R.~C.~Hou and Y.~Matsuo,
%  ``Lie 3-Algebra and Multiple M2-branes,''
  arXiv:0804.2110 [hep-th].
  %%CITATION = ARXIV:0804.2110;%%

%\cite{Gran:2008vi}
\bibitem{Gran:2008vi}
  U.~Gran, B.~E.~W.~Nilsson and C.~Petersson,
%  ``On relating multiple M2 and D2-branes,''
  arXiv:0804.1784 [hep-th].
  %%CITATION = ARXIV:0804.1784;%%

%\cite{Distler:2008mk}
\bibitem{Distler:2008mk}
  J.~Distler, S.~Mukhi, C.~Papageorgakis and M.~Van Raamsdonk,
  %``M2-branes on M-folds,''
  JHEP {\bf 0805}, 038 (2008)
  [arXiv:0804.1256 [hep-th]].
  %%CITATION = JHEPA,0805,038;%%

%\cite{Lambert:2008et}
\bibitem{Lambert:2008et}
  N.~Lambert and D.~Tong,
%  ``Membranes on an Orbifold,''
  arXiv:0804.1114 [hep-th].
  %%CITATION = ARXIV:0804.1114;%%

%\cite{Morozov:2008cb}
\bibitem{Morozov:2008cb}
  A.~Morozov,
  %``On the Problem of Multiple M2 Branes,''
  JHEP {\bf 0805}, 076 (2008)
  [arXiv:0804.0913 [hep-th]].
  %%CITATION = JHEPA,0805,076;%%

%\cite{VanRaamsdonk:2008ft}
\bibitem{VanRaamsdonk:2008ft}
  M.~Van Raamsdonk,
  %``Comments on the Bagger-Lambert theory and multiple M2-branes,''
  JHEP {\bf 0805}, 105 (2008)
  [arXiv:0803.3803 [hep-th]].
  %%CITATION = JHEPA,0805,105;%%

%\cite{Berman:2008be}
\bibitem{Berman:2008be}
  D.~S.~Berman, L.~C.~Tadrowski and D.~C.~Thompson,
%  ``Aspects of Multiple Membranes,''
  arXiv:0803.3611 [hep-th].
  %%CITATION = ARXIV:0803.3611;%%

%\cite{Bandres:2008vf}
\bibitem{Bandres:2008vf}
  M.~A.~Bandres, A.~E.~Lipstein and J.~H.~Schwarz,
%  ``N = 8 Superconformal Chern--Simons Theories,''
  JHEP {\bf 0805}, 025 (2008)
  [arXiv:0803.3242 [hep-th]].
  %%CITATION = JHEPA,0805,025;%%

%\cite{Mukhi:2008ux}
\bibitem{MP}
  S.~Mukhi and C.~Papageorgakis,
  %``M2 to D2,''
  JHEP {\bf 0805}, 085 (2008)
  [arXiv:0803.3218 [hep-th]].
  %%CITATION = JHEPA,0805,085;%%

%%%%%%%%%%%%%%%%%%%%%%%%%%%%%%%%%%%%%%%%%%%%%%%%%%%%%%%%%%%%%%%%%%%%%%%%%%%%%%
%%%%%%%%%%%%%%%%%%%%%%%%%%%%%%%%%%%%%%%%%%%%%%%%%%%%%%%%%%%%%%%%%%%%%%%%%%%%%%

%\cite{Freedman:1999gp}
\bibitem{FGPW}
  D.~Z.~Freedman, S.~S.~Gubser, K.~Pilch and N.~P.~Warner,
%  ``Renormalization group flows from holography supersymmetry and a
%  c-theorem,''
  Adv.\ Theor.\ Math.\ Phys.\  {\bf 3}, 363 (1999)
  [arXiv:hep-th/9904017].
  %%CITATION = 00203,3,363;%%

%\cite{Leigh:1995ep}
\bibitem{LS}
  R.~G.~Leigh and M.~J.~Strassler,
%  ``Exactly Marginal Operators And Duality In Four-Dimensional N=1
%  Supersymmetric Gauge Theory,''
  Nucl.\ Phys.\  B {\bf 447}, 95 (1995)
  [arXiv:hep-th/9503121].
  %%CITATION = NUPHA,B447,95;%%

%\cite{Nemeschansky:2004yh}
\bibitem{NW04}
  D.~Nemeschansky and N.~P.~Warner,
%  ``A family of M-theory flows with four supersymmetries,''
  arXiv:hep-th/0403006;
  %%CITATION = HEP-TH/0403006;%%
%
%\cite{Gowdigere:2003jf}
%\bibitem{GNW}
  C.~N.~Gowdigere, D.~Nemeschansky and N.~P.~Warner,
%  ``Supersymmetric solutions with fluxes from algebraic Killing spinors,''
  Adv.\ Theor.\ Math.\ Phys.\  {\bf 7}, 787 (2004)
  [arXiv:hep-th/0306097];
  %%CITATION = 00203,7,787;%%
%
%\cite{Gowdigere:2002uk}
%\bibitem{GW}
  C.~N.~Gowdigere and N.~P.~Warner,
%  ``Flowing with eight supersymmetries in M-theory and F-theory,''
  JHEP {\bf 0312}, 048 (2003)
  [arXiv:hep-th/0212190].
  %%CITATION = JHEPA,0312,048;%%

%\cite{Hosomichi:2008qk}
\bibitem{HLL}
  K.~Hosomichi, K.~M.~Lee and S.~Lee,
%  ``Mass-Deformed Bagger-Lambert Theory and its BPS Objects,''
  arXiv:0804.2519 [hep-th].
  %%CITATION = ARXIV:0804.2519;%%

%\cite{Gomis:2008cv}
\bibitem{GSP}
  J.~Gomis, A.~J.~Salim and F.~Passerini,
%  ``Matrix Theory of Type IIB Plane Wave from Membranes,''
  arXiv:0804.2186 [hep-th].
  %%CITATION = ARXIV:0804.2186;%%

%\cite{Cremmer:1978km}
\bibitem{CJS}
  E.~Cremmer, B.~Julia and J.~Scherk,
%  ``Supergravity theory in 11 dimensions,''
  Phys.\ Lett.\  B {\bf 76}, 409 (1978);
  %%CITATION = PHLTA,B76,409;%%
%
%\cite{Cremmer:1979up}
%\bibitem{CJ}
  E.~Cremmer and B.~Julia,
%  ``The SO(8) Supergravity,''
  Nucl.\ Phys.\  B {\bf 159}, 141 (1979).
  %%CITATION = NUPHA,B159,141;%%

%\cite{de Wit:1981eq}
\bibitem{dN82}
  B.~de Wit and H.~Nicolai,
%  ``N=8 Supergravity With Local SO(8) X SU(8) Invariance,''
  Phys.\ Lett.\  B {\bf 108}, 285 (1982);
  %%CITATION = PHLTA,B108,285;%%
%
%\cite{de Wit:1982ig}
%\bibitem{dN82-1}
  B.~de Wit and H.~Nicolai,
%  ``N=8 Supergravity,''
  Nucl.\ Phys.\  B {\bf 208}, 323 (1982).
  %%CITATION = NUPHA,B208,323;%%

%\cite{Cremmer:1978ds}
\bibitem{CJ78}
  E.~Cremmer and B.~Julia,
%  ``The N=8 Supergravity Theory. 1. The Lagrangian,''
  Phys.\ Lett.\  B {\bf 80}, 48 (1978).
  %%CITATION = PHLTA,B80,48;%%

%\cite{Nicolai:1985hs}
\bibitem{NW}
  H.~Nicolai and N.~P.~Warner,
%  ``The SU(3) X U(1) Invariant Breaking Of Gauged N=8 Supergravity,''
  Nucl.\ Phys.\  B {\bf 259}, 412 (1985).
  %%CITATION = NUPHA,B259,412;%%

%\cite{Warner:1983vz}
\bibitem{Warner83}
  N.~P.~Warner,
%  ``Some New Extrema Of The Scalar Potential Of Gauged N=8 Supergravity,''
  Phys.\ Lett.\  B {\bf 128}, 169 (1983),
  %%CITATION = PHLTA,B128,169;%%
%
%\cite{Warner:1983du}
%\bibitem{Warner84}
%  N.~P.~Warner,
%  ``Some Properties Of The Scalar Potential In Gauged Supergravity Theories,''
  Nucl.\ Phys.\  B {\bf 231}, 250 (1984).
  %%CITATION = NUPHA,B231,250;%%

%\cite{Benna:2008zy}
\bibitem{BKKS}
  M.~Benna, I.~Klebanov, T.~Klose and M.~Smedback,
  %``Superconformal Chern-Simons Theories and AdS_4/CFT_3 Correspondence,''
  arXiv:0806.1519 [hep-th].
  %%CITATION = ARXIV:0806.1519;%%

%\cite{Strassler:1998iz}
\bibitem{Strassler98}
  M.~J.~Strassler,
%  ``On renormalization group flows and exactly marginal operators in three
%  dimensions,''
  arXiv:hep-th/9810223.
  %%CITATION = HEP-TH/9810223;%%

%\cite{Merlatti:2000ed}
\bibitem{Merlatti}
  P.~Merlatti,
%  ``M-theory on AdS(4) x Q(111): The complete Osp(2|4) x SU(2) x SU(2) x  SU(2)
%  spectrum from harmonic analysis,''
  Class.\ Quant.\ Grav.\  {\bf 18}, 2797 (2001)
  [arXiv:hep-th/0012159];
  %%CITATION = CQGRD,18,2797;%%
%
%\cite{Ceresole:1999zg}
%\bibitem{CDDF}
  A.~Ceresole, G.~Dall'Agata, R.~D'Auria and S.~Ferrara,
%  ``M-theory on the Stiefel manifold and 3d conformal field theories,''
  JHEP {\bf 0003}, 011 (2000)
  [arXiv:hep-th/9912107];
  %%CITATION = JHEPA,0003,011;%%
%
%\cite{Fabbri:1999ay}
%\bibitem{FFGT1}
  D.~Fabbri, P.~Fre, L.~Gualtieri and P.~Termonia,
%  ``Osp(N|4) supermultiplets as conformal superfields on d(AdS(4)) and the
%  generic form of N = 2, D = 3 gauge theories,''
  Class.\ Quant.\ Grav.\  {\bf 17}, 55 (2000)
  [arXiv:hep-th/9905134].
  %%CITATION = CQGRD,17,55;%%

%\cite{Ceresole:1984hr}
\bibitem{CFN}
  A.~Ceresole, P.~Fre and H.~Nicolai,
%  ``Multiplet Structure And Spectra Of N=2 Supersymmetric Compactifications,''
  Class.\ Quant.\ Grav.\  {\bf 2}, 133 (1985).
  %%CITATION = CQGRD,2,133;%%

%\cite{Ahn:1998sv}
\bibitem{AOT}
  C.~Ahn, K.~Oh and R.~Tatar,
%  ``Branes, orbifolds and the three dimensional N = 2 SCFT in the large N
%  limit,''
  JHEP {\bf 9811}, 024 (1998)
  [arXiv:hep-th/9806041].
  %%CITATION = JHEPA,9811,024;%%

%\cite{Ahn:1999ec}
\bibitem{Ahn99}
  C.~Ahn,
%  ``N = 2 SCFT and M theory on AdS(4) x Q(1,1,1),''
  Phys.\ Lett.\  B {\bf 466}, 171 (1999)
  [arXiv:hep-th/9908162].
  %%CITATION = PHLTA,B466,171;%%

%\cite{Fabbri:1999hw}
\bibitem{FFGRTZZ}
  D.~Fabbri, P.~Fre', L.~Gualtieri, C.~Reina, A.~Tomasiello, A.~Zaffaroni and A.~Zampa,
%  ``3D superconformal theories from Sasakian seven-manifolds: New  nontrivial
%  evidences for AdS(4)/CFT(3),''
  Nucl.\ Phys.\  B {\bf 577}, 547 (2000)
  [arXiv:hep-th/9907219].
  %%CITATION = NUPHA,B577,547;%%

%\cite{Fabbri:1999mk}
\bibitem{FFGT}
  D.~Fabbri, P.~Fre, L.~Gualtieri and P.~Termonia,
%  ``M-theory on AdS(4) x M(111): The complete Osp(2|4) x SU(3) x SU(2)
%  spectrum from harmonic analysis,''
  Nucl.\ Phys.\  B {\bf 560}, 617 (1999)
  [arXiv:hep-th/9903036].
  %%CITATION = NUPHA,B560,617;%%

%\cite{Ahn:2002xr}
\bibitem{Ahn02-3}
  C.~Ahn,
%  ``More Penrose limit of AdS(4) x N(0,1,0) and N = 3 gauge theory,''
  Mod.\ Phys.\ Lett.\  A {\bf 17}, 1847 (2002)
  [arXiv:hep-th/0206176],
  %%CITATION = MPLAE,A17,1847;%%
%
%\cite{Ahn:2002sb}
%\bibitem{Ahn02-2}
%  C.~Ahn,
%  ``Penrose limit of AdS(4) x V(5,2) and operators with large R charge,''
  Mod.\ Phys.\ Lett.\  A {\bf 17}, 2067 (2002)
  [arXiv:hep-th/0206029],
  %%CITATION = MPLAE,A17,2067;%%
%
%\cite{Ahn:2002nt}
%\bibitem{Ahn02-1}
%  C.~Ahn,
%  ``Comments on Penrose limit of AdS(4) x M(1,1,1),''
  Phys.\ Lett.\  B {\bf 540}, 111 (2002)
  [arXiv:hep-th/0205109],
  %%CITATION = PHLTA,B540,111;%%
%
%\cite{Ahn:2002qj}
%\bibitem{Ahn02}
%  C.~Ahn,
%  ``More on Penrose limit of AdS(4) x Q(1,1,1),''
  Phys.\ Lett.\  B {\bf 539}, 281 (2002)
  [arXiv:hep-th/0205008].
  %%CITATION = PHLTA,B539,281;%%

%\cite{Ahn:2002eh}
\bibitem{AT02-1}
  C.~Ahn and T.~Itoh,
%  ``The 11-dimensional metric for AdS/CFT RG flows with common SU(3)
%  invariance,''
  Nucl.\ Phys.\  B {\bf 646}, 257 (2002)
  [arXiv:hep-th/0208137],
  %%CITATION = NUPHA,B646,257;%%
%\cite{Ahn:2001kw}
%\bibitem{AT02}
%  C.~Ahn and T.~Itoh,
%  ``An N = 1 supersymmetric G(2)-invariant flow in M-theory,''
  Nucl.\ Phys.\  B {\bf 627}, 45 (2002)
  [arXiv:hep-th/0112010],
  %%CITATION = NUPHA,B627,45;%%
%\cite{Ahn:2001nw}
%\bibitem{AI}
%  C.~Ahn and T.~Itoh,
%  ``Dielectric branes in non-supersymmetric SO(3)-invariant perturbation of
%  three-dimensional N = 8 Yang-Mills theory,''
  Phys.\ Rev.\  D {\bf 64}, 086006 (2001)
  [arXiv:hep-th/0105044];
  %%CITATION = PHRVA,D64,086006;%%
%
%\cite{Ahn:2002qga}
%\bibitem{AW03}
  C.~Ahn and K.~Woo,
%  ``Domain wall from gauged d = 4, N = 8 supergravity. II,''
  JHEP {\bf 0311}, 014 (2003)
  [arXiv:hep-th/0209128],
  %%CITATION = JHEPA,0311,014;%%
%
%\cite{Ahn:2001by}
%\bibitem{AW02}
%  C.~Ahn and K.~Woo,
%  ``Domain wall and membrane flow from other gauged d = 4, n = 8  supergravity.
%  I,''
  Nucl.\ Phys.\  B {\bf 634}, 141 (2002)
  [arXiv:hep-th/0109010];
  %%CITATION = NUPHA,B634,141;%%
%
%\cite{Ahn:1999zy}
%\bibitem{AR99}
  C.~Ahn and S.~J.~Rey,
%  ``More CFTs and RG flows from deforming M2/M5-brane horizon,''
  Nucl.\ Phys.\  B {\bf 572}, 188 (2000)
  [arXiv:hep-th/9911199],
  %%CITATION = NUPHA,B572,188;%%
%
%\cite{Ahn:1999dq}
%\bibitem{AR99-1}
%  C.~Ahn and S.~J.~Rey,
%  ``Three-dimensional CFTs and RG flow from squashing M2-brane horizon,''
  Nucl.\ Phys.\  B {\bf 565}, 210 (2000)
  [arXiv:hep-th/9908110].
  %%CITATION = NUPHA,B565,210;%%

\end{thebibliography}
\end{document}